\documentclass{IEEEtran}
\usepackage{amssymb}
\usepackage{amsfonts}
\usepackage{amsmath}
\usepackage{algorithm}
\usepackage{algorithmic}
\usepackage{tikz}
\usepackage{mathtools}
\usepackage{bm}
\usepackage{graphicx,subfigure,cite,multicol,multirow,diagbox,booktabs,array}
\usepackage{color}
\usetikzlibrary{shapes.geometric, arrows}
\ifCLASSINFOpdf
\else
\fi

\begin{document}
\title{Secure Multigroup Multicast Communication Systems via Intelligent Reflecting Surface}
\author{Weiping~Shi, Jiayu~Li, Guiyang~Xia, Yuntian~Wang, \\Xiaobo~Zhou, Yonghui~Zhang, Feng~Shu, \emph{Member},~\emph{IEEE}
\thanks{W. Shi, J. Li, G. Xia, Y. Wang and F. Shu are with School of Electronic and Optical Engineering, Nanjing University of Science and Technology, Nanjing, 210094, China.}
\thanks{X. Zhou is with the School of Physics and Electronic Engineering, Fuyang 
Normal University, Fuyang 236037, China, zxb@fynu.edu.cn}
\thanks{Y. Zhang and F. Shu are  with the School of Information and Communication Engineering, Hainan University,~Haikou,~570228, China.}
}
\maketitle
\begin{abstract}
This paper considers a secure multigroup multicast multiple-input single-output (MISO) communication system aided by an intelligent reflecting surface (IRS). Specifically, we aim to minimize the transmit power at the Alice via jointly optimizing the transmit beamformer, AN vector and phase shifts at the IRS subject to the secrecy rate constraints as well as the unit modulus constraints of IRS phase shifts. However, the optimization problem is non-convex and directly solving it is intractable. To tackle the optimization problem, we first transform it into a semidefinite relaxation (SDR) problem, and then alternately update the transmit beamformer and AN matrix as well as the phase shifts at the IRS. In order to reduce the high computational complexity, we further propose a low-complexity algorithm based on second-order cone programming (SOCP). We decouple the optimization problem into two sub-problems and optimize the transmit beamformer, AN vector and the phase shifts alternately by solving two corresponding SOCP sub-problem. Simulation results show that the proposed SDR and SOCP schemes require half or less transmit power than the scheme without IRS, which demonstrates the advantages of introducing IRS and the effectiveness of the proposed methods.

\end{abstract}



\IEEEpeerreviewmaketitle
\section{Introduction}
For beyond fifth-generation (5G) and sixth-generation (6G) communication systems, due to the fact that a massive number of mobile users are required to be supported, various techniques were proposed to improve the spectrum efficiency and energy efficiency, such as massive multiple-input multiple-output (MIMO), millimeter wave (mmWave) communication and so on\cite{huang1,wuyongpeng,QinYL_DOA_MIMO}. However, the hardware cost and energy consumption of these technologies increase as the number of deployed base stations (BSs) increase.  Moreover, excessive number of active components can also lead to serious interference issue in wireless
networks. A new and revolutionary technology, intelligent reflecting surface (IRS) achieves high spectrum efficiency and energy efficiency with low hardware cost \cite{ZhongCJ_IRS}. Specifically, an IRS consists of a large number of low-cost, passive and reflecting units, which reflect the signal by dynamically adjusting the phase shifts of the elements. The reflected signals gather at the desired receiver to improve the received signal strength, while destructively at the non-intended receiver for reducing the interference \cite{IRS_mag}.

Due to those above-mentioned benefits, several IRS-aided wireless communication systems were investigated to enhance the communication performance, such as the received signal-to-noise ratio (SNR), energy efficiency and  secrecy rate (SR).
Explicity, the authors in \cite{wu_one} proposed a joint active and passive beamforming design for an IRS-assisted single-user and multiple-user multiple-input single-output (MISO) communication system, where semidefinite relaxation (SDR) and alternating optimization algorithm are performed to optimize the transmit beamformer and the phase shifts at the IRS for minimizing the total transmit power.
In order to reduce the computational complexity incurred by SDR technique, Yu $\emph{et~al}.$  proposed a pair of efficient algorithms (i.e., fixed point iteration and manifold techniques) to optimize the phase shifts at the IRS. It is worth mentioning that the two proposed methods are capable of obtaining locally optimal solutions \cite{manifold}.
To maximize the energy efficiency in MISO systems, the authors of \cite{Energy_Efficiency} performed gradient descent/sequential fractional programming method to optimize the phase shifts at the IRS, and using Dinkelbach¡¯s method for optimizing the power allocation factor.

On the other hand, the broadcast nature of wireless channels leads to that the information sent to a legitimate receiver can be also gathered by the unintended receivers (eavesdroppers) \cite{chenXM_secury,wangMH_secury,RFDA-HJS,XiaGY_Green_sec}. 
Traditionally, the security problem was ensured by the encryption technique, which requires complex key management. By contrast, physical layer security provides a new approach by fully exploring the random nature of communication channels to arrive a secure transmission \cite{zhangN_secury,zhou_UAV,XiaGY_TWC_sec}. As such, the complex key management is circumvented. It is known that the IRS could enhance the received signal power to the legitimate receiver, while the power received at eavesdroppers (Eves) will be also enhanced. Therefore, how to improve the secrecy performance for the IRS-aided communication systems becomes a non-negligible problem \cite{SHEN_Secrecy_IRS, XU_AN_MU_sec, PAN_SEC_MIMO}.
Shen $\emph{et~al}.$ considered the secrecy rate (SR) maximization problem in a secure single-user MISO communication with an IRS, in which  Majorization$-$Minimization (MM) algorithm was applied and a closed-form solution was obtained \cite{SHEN_Secrecy_IRS}.
Resource allocation problem for secure IRS-aided multiuser MISO system and SR maximization problem for secure MIMO system were investigated by jointly optimizing the transmit beamformer, AN covariance matrix and the phase shifts at the IRS, respectively \cite{XU_AN_MU_sec,PAN_SEC_MIMO}. However, the authors in \cite{XU_AN_MU_sec} did not consider the links between BS and users.
In addition, secure simultaneous wireless information and power transfer (SWIPT) system assisted by the IRS was optimized to maximize the harvested power in \cite{shi_SWIPT}. All the mentioned systems are demonstrated that the significantly performance gains (e.g., security or the received signal power) of the communication system can be achieved by introducing an IRS.
In practical communications, obtaining an accurate channel state information (CSI) of communication nodes is challenging, especially for the case that an IRS is further considered in the communication system. To solve this issue, several channel estimation methods and robust transmission designs were proposed\cite{CSI_MIMO,CSI_OFDM,huang2}.
The authors designed the transmission
protocol and verified the efficiency of the channel estimation based on the ON/OFF model \cite{CSI_MIMO,CSI_OFDM}. In Addition, Huang $\emph{et~al}.$ proposed a deep learning method to promptly optimize the transmit beamformer and the phase shifts at the IRS \cite{huang2}.

Clearly, the above-mentioned contributions focus on exploiting IRS for enhancing the performance profits in unicast or broadcast transmissions.
For broadcast transmissions, BS sends the same stream to all sheathed users, which can not provide personalized service timely according to the requirement of each customer. For unicast transmissions, BS sends an independent data stream to each user, which causes severe interference and high system complexity in the fact of a large number of users. 
To address these issues, the multicast transmission has attracted widely attention and it has great potential in many applications such as popular TV programme and live video streaming\cite{MassiveMIMO_Mulcast}. Physical layer multicasting via beamforming is shown to be useful for alleviating the pressure of huge wireless data traffic and for boosting the spectrum and energy efficiency \cite{ADMM_MUG,XU_MU_G,qichenhao_Mulcast}. An IRS-assisted multigroup multicast MISO communication system is studied to maximize the sum rate of all the multicasting groups by the joint optimization of the precoding matrix at the BS and the reflection coefficients at the IRS\cite{IRS_MUgMUc}. The authors proposed two  efficient algorithms based on the MM algorithm framework. Since the transmitter serves multiple legitimate users by using single beamforming vector, the multicast system becomes more vulnerable to eavesdropping from the perspective of security.  Therefore, it is necessary to investigate the performance advantages when an IRS is considered in secure multigroup multicast systems.

Motivated by the above discussions, we investigate the physical layer security  for an IRS-aided multigroup multicast system. To elaborate, the multiple-antenna Alice transmits independently confidential information data stream to each legitimate multiple group in the presence of Eves. To improve the security of the system, Alice also transmits AN to disturb Eves' decoding. All the legitimate receivers and Eves are assumed to be equipped with a single antenna. The legitimate users in the same group receive the same information, but they are interfered by the signals sent to other groups. The contributions of our work are summarized as follows:
\begin{enumerate}
  \item
  For the first time we formulate a transmit power minimization problem by jointly optimizing the transmit beamforming vector and the AN vector of the Alice as well as the phase shifts at the IRS for the IRS-aided multigroup multicast system in the presence of Eves subject to a non-convex uni-modular constraint and the SR constraint.
  \item
  To solve the problem, an iterative and alternating optimization algorithm based on SDR technique is proposed. Specifically, we first transform the optimization problem into an SDR problem by dropping the rank-one constraints. Then the alternating optimization method is applied to separately optimize the transmit beamforming matrix and the AN matrix as well as the phase shifts at the IRS. In each subproblem, MM algorithm is used to obtain the upper bound of the concave logarithm function. Thus each subproblem can be transformed into a convex problem and then be solved directly. Furthermore, the Gaussian randomization method is adopted to obtain a high quality sub-optimal solution.
  \item
  To reduce the computational complexity of the SDR algorithm, we further propose an efficient algorithm in an iterative manner based on second-order cone programming (SOCP) technique. Specifically, we decouple the problem into two subproblems for optimizing the transmit beamforming vector, AN vector and the phase shift at the IRS, respectively. We handle the non-convex SR constraints by introducing the first-order Taylor expansion and then transform them into a SOCP for each subproblem, which is finally solved by successive convex approximation (SCA) method.
  Simulation results demonstrate that our proposed two IRS-aided schemes are capable of saving transmit power significantly, when compared with no IRS-aided scheme.
\end{enumerate}

This rest of paper is organized as follows. Section II describes an IRS-aided system model for a secure multigroup multicast MISO communication system. Then, an associated power minimization problem is formulated. The SDR-based method is developed to solve the non-convex optimization problem in Section III. Section IV  provides an SOCP-based alternative method to reduce the computational complexity of the SDR approach. Section V shows our numerical simulation results to validate the performance improvement of the proposed algorithms. Finally, Section VI draws our conclusions.

\emph{Notations}: Scalars are presented by lowercase letters. Vectors and matrices are denoted by  boldface uppercase and lowercase letters, respectively. $|\cdot|$ denotes the modulus of a scalar and $\|\cdot\|$ denotes the Euclidean norm of a vector. $(\cdot)^H$, $\mathrm{Tr}(\cdot)$ and $\angle(\cdot)$ denote the conjugate transpose, the trace of a matrix and the angle of a complex number, respectively.

\section{System Model and Problem Formulation}
\begin{figure}[htb]
  \centering
  \includegraphics[width=0.48\textwidth]{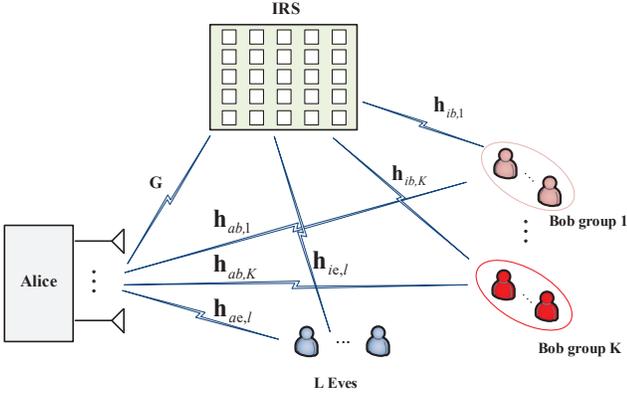}
  \caption{An IRS-aided secure multigroup multicast communication system.}\label{Sys_Mod}
  \label{sys}
\end{figure}

Fig.~\ref{Sys_Mod} sketches an IRS-aided downlink multigroup multicast MISO system, where Alice is equipped with $M$ transmit antennas serving multiple users (Bobs) in $K$ multicast groups with assistance of an IRS in the presence of $L$ Eves. All receivers are equipped with a single antenna while the IRS has $N$ reflecting elements. Denoting $\mathcal{G}_k (\forall k\in \mathcal{K}=\{1,...,K\})$ as the $k$-th group of the desired users, and the total number of user is denoted as $T = \sum_{k=1}^{K}|\mathcal{G}_k|$. Assume that each user belongs to one unique multicast group, i.e., $\mathcal{G}_i\bigcap\mathcal{G}_j=\emptyset, i\neq j, \forall i,j \in \mathcal{K}$. Moreover, the set of Eves is denoted as $\mathcal{L}=\{1,...,L\}$. The transmitted signal at Alice can be expressed as
\begin{equation}\label{Tx signal s}
\mathbf{x}=\sum_{k=1}^{K}\mathbf{w}_k s_k+\mathbf{q}_{AN},
\end{equation}
where $\mathbf{w}_k\in\mathbb{C}^{M\times 1}$ and $\mathbf{q}_{AN}\in\mathbb{C}^{M\times 1}$ denote  the transmit beamforming vector which forces the confidential message (CM) to the desired $k$-th group Bobs and AN to disturb Eves for enhancing physical layer security, respectively. In addition, $s_k$ denotes the CM for Bobs in the $k$-th group .
Without loss of generality, we assume $s_k$ is independently and identically distributed (i.i.d.) circularly symmetric complex Gaussian (CSCG) random variable with zero mean and unit variance, i.e., $s_k\sim\mathcal{C}\mathcal{N}(0,1)$. Thus, the total transmit power at the Alice is
$\mathbb{E}\left\{\mathbf{x}^H\mathbf{x}\right\}=\sum_{i=k}^{K}\|\mathbf{w}_k\|^2+\|\mathbf{q}_{AN}\|^2$

In this paper, a quasi-static fading environment is assumed. The baseband equivalent channel responses from
 Alice to the IRS, from Alice to the $j$-th user in the $k$-th Bob group, from the Alice to the $l$-th Eve, from the IRS to the $j$-th user in the $k$-th Bob group and from the IRS to the $l$-th Eve are denoted by $\mathbf{G}\in\mathbb{C}^{N\times M}$, $\mathbf{h}_{ab,kj}^H\in\mathbb{C}^{1\times M}$, $\mathbf{h}_{ae,l}^H\in\mathbb{C}^{1\times M}$, $\mathbf{h}_{ib,kj}^H\in\mathbb{C}^{1\times N}$ and $\mathbf{h}_{ie,l}^H\in\mathbb{C}^{1\times N}$, respectively. The reflecting coefficient channel of the IRS is denoted as $\mathbf{\Theta}=\mathrm{diag}(\mathbf{\beta}\mathbf{e}^{j\mathbf{\theta}_{1}},\cdots,\mathbf{\beta}\mathbf{e}^{j\mathbf{\theta}_{N}})$, where $\beta\in [0,1]$ denotes the amplitude reflection coefficient and $\bm{\theta}=[\theta_1,\cdots,\theta_N]$ is the phase shift vector at the IRS, respectively \cite{Lijiayu_IRS}. For simplification, $\beta = 1$ is assumed. As such, the received signal at the $j$-th user in the $k$-th Bob group can be written as
\begin{align}\label{Rx_signal yb}
&y_{b,kj}
=(\mathbf{h}_{ib,kj}^H\mathbf{\Theta}\mathbf{G}+\mathbf{h}_{ab,kj}^H)(\sum_{k=1}^{K}\mathbf{w}_k s_k+\mathbf{q}_{AN})+n_{bk},
\end{align}
where $n_{bk}\sim\mathcal{C}\mathcal{N}(0,\sigma_b^2)$ is the complex additive white Gaussian noise (AWGN). Similarly, the received signal at the $l$-th Eve is
\begin{align}\label{Rx_signal ye}
y_{e,l}=(\mathbf{h}_{ie,l}^H\mathbf{\Theta}\mathbf{G}+\mathbf{h}_{ae,l}^H)(\sum_{k=1}^{K}\mathbf{w}_k s_k+\mathbf{q}_{AN})+n_{el},
\end{align}
where $n_{el}$ is the complex AWGN variable following the distribution $n_{el}\sim\mathcal{C}\mathcal{N}(0,\sigma_e^2)$. Moreover, we assume that $\sigma_{b}^2=\sigma_{e}^2=\sigma^2$. According to (\ref{Rx_signal yb}) and (\ref{Rx_signal ye}), the achievable transmission rate at the $j$-th user in the $k$-th Bob group and the $l$-th Eve intending to wiretap the $j$-th legitimate user in the $k$-th Bob group can be expressed as\cite{cui_security}
\begin{align}\label{Rb}
R_{b,kj}=\log_2\left(1+\frac{|(\mathbf{h}_{ib,kj}^H\mathbf{\Theta}\mathbf{G}+\mathbf{h}_{ab,kj}^H)\mathbf{w}_k|^2}
{\splitfrac{\sum_{g\neq k}^{K}|(\mathbf{h}_{ib,kj}^H\mathbf{\Theta}\mathbf{G}+\mathbf{h}_{ab,kj}^H)\mathbf{w}_g|^2}
{+|(\mathbf{h}_{ib,kj}^H\mathbf{\Theta}\mathbf{G}+\mathbf{h}_{ab,kj}^H)\mathbf{q}_{AN}|^2+\sigma^2}}\right)
\end{align}
and
\begin{align}\label{Re}
R_{e,l}
=\log_2\left(1+\frac{|(\mathbf{h}_{ie,l}^H\mathbf{\Theta}\mathbf{G}+\mathbf{h}_{ae,l}^H)\mathbf{w}_k|^2}
{\splitfrac{\sum_{g\neq k}^{K}|(\mathbf{h}_{ie,l}^H\mathbf{\Theta}\mathbf{G}+\mathbf{h}_{ae,l}^H)\mathbf{w}_g|^2}
{+|(\mathbf{h}_{ie,l}^H\mathbf{\Theta}\mathbf{G}+\mathbf{h}_{ae,l}^H)\mathbf{q}_{AN}|^2+\sigma^2}}\right),
\end{align}
respectively. The corresponding achievable SR of user $j$ in Bob group $k$ is defined by \cite{wuXM_Secury}
\begin{align}\label{Rs}
R_{s,kj}=\big[0,\min_{j\in \mathcal{G}_k}R_{b,kj}- \max_{\forall l}R_{e,l}\big]^+\nonumber\\
=\big[0,\min_{j\in \mathcal{G}_k,\forall l}(R_{b,kj}- {}R_{e,l})\big]^+.
\end{align}

In this paper, we aim to minimize the total transmit power required at the Alice subject to the minimum SR constraints at Bobs and phase shift constraints
at the IRS by jointly optimizing the transmit beamformings, AN at the Alice and the reflect phase shifts at the IRS. Let us define $\mathbf{v}=[\mathbf{e}^{j\theta_{1}},\cdots,\mathbf{e}^{j\theta_{N}}]^H\in\mathbb{C}^{N\times 1}$,
$\mathbf{u}=[\mathbf{v},1]$,
$\mathbf{H}_{b,kj}=\mathrm{diag}\{\mathbf{h}_{ib,kj}^H\}\mathbf{G} \in\mathbb{C}^{N\times M}$, $\mathbf{H}_{e,l}=\mathrm{diag}\{\mathbf{h}_{ie,l}^H\}\mathbf{G}\in\mathbb{C}^{N\times M}$, $\mathbf{H}_{kj} = [\mathbf{H}_{b,kj};\mathbf{h}_{ab,kj}^H]$, and
$\mathbf{H}_{l} = [\mathbf{H}_{e,l};\mathbf{h}_{ae,l}^H]$. As such, $R_{b,kj}$ and $R_{e,l}$ can be respectively rewritten as (\ref{Rb_1}) and (\ref{Re_1}) at the top of the next page.
\newcounter{mytempeqncnt}
\begin{figure*}[!t]
\normalsize
\setcounter{mytempeqncnt}{\value{equation}}
\setcounter{equation}{6}
\begin{equation}
\begin{aligned}\label{Rb_1}
R_{b,kj}=\log_2\left(1+\frac{|(\mathbf{v}^H\mathbf{H}_{b,kj}+\mathbf{h}_{ab,kj}^H)\mathbf{w}_k|^2}
{\sum_{g\neq k}^{K}|(\mathbf{v}^H\mathbf{H}_{b,kj}+\mathbf{h}_{ab,kj}^H)\mathbf{w}_g|^2
+|(\mathbf{v}^H\mathbf{H}_{b,kj}+\mathbf{h}_{ab,kj}^H)\mathbf{q}_{AN}|^2+\sigma^2}\right)\\
=\log_2\left(1+\frac{|\mathbf{u}^H\mathbf{H}_{kj}\mathbf{w}_k|^2}
{\sum_{g\neq k}^{K}|\mathbf{u}^H\mathbf{H}_{kj}\mathbf{w}_g|^2
+|\mathbf{u}^H\mathbf{H}_{kj}\mathbf{q}_{AN}|^2+\sigma^2}\right)
=\log_2(1+\mathrm{SINR}_{b(k,j)}),~~\forall k,j,\\
\end{aligned}
\end{equation}
\begin{equation}
\begin{aligned}\label{Re_1}
R_{e,l}
=\log_2\left(1+\frac{|(\mathbf{v}^H\mathbf{H}_{e,l}+\mathbf{h}_{ae,l}^H)\mathbf{w}_k|^2}
{\sum_{g\neq k}^{K}|(\mathbf{v}^H\mathbf{H}_{e,l}+\mathbf{h}_{ae,l}^H)\mathbf{w}_g|^2+|(\mathbf{v}^H\mathbf{H}_{e,l}
+\mathbf{h}_{ae,l}^H)\mathbf{q}_{AN}|^2+\sigma^2}\right)\\
=\log_2\left(1+\frac{|\mathbf{u}^H\mathbf{H}_{l}\mathbf{w}_k|^2}
{\sum_{g\neq k}^{K}|\mathbf{u}^H\mathbf{H}_{l}\mathbf{w}_g|^2+|\mathbf{u}^H\mathbf{H}_{l}
\mathbf{q}_{AN}|^2+\sigma^2}\right)
=\log_2(1+\mathrm{SINR}_{e(k,l)}),~~\forall k,l.
\end{aligned}
\end{equation}
\setcounter{equation}{8}
\hrulefill
\end{figure*}
Thus, the resultant optimization problem can be mathematically formulated as
\begin{subequations}\label{P1}
\begin{align}
\text{(P1):}&\min_{\{\mathbf{w}_k\},\mathbf{q}_{AN},\mathbf{v}} \sum_{k=1}^{K}\|\mathbf{w}_k\|^2+\|\mathbf{q}_{AN}\|^2\\
&~~\text{s. t.}~~ R_{s,kj} \geq \gamma_s,~~\forall k \in \mathcal{K}, l \in \mathcal{L}, j\in \mathcal{G}_k,\label{Rs_const}\\
&~~~~~~~~|v_n|^2 = 1,~~\forall n=1\cdots N,\label{vnunit}
\end{align}
\end{subequations}
where $\gamma_s \geq 0$ is the minimum target SR for Bobs. However, it is observed that problem (P1) is a non-convex optimization problem and thus difficult to solve optimally. This is due to the fact that constraints (\ref{Rs_const}) are non-convex with respect to \{$\mathbf{w}_k$, $\mathbf{q}_{AN}$, $\mathbf{v}$\} and (\ref{vnunit}) are non-linear equality constraints. As a result, we will propose two efficient algorithms to solve problem (P1) sub-optimally.
\section{SDR-based Alternating Optimization Method}
In this section, we propose an SDR-based  alternating optimization (AO) method to solve problem (P1). To facilitate processing, problem (P1) is first relaxed as an SDR problem and then the SDR problem is decoupled into two subproblems. For each subproblem, we apply MM algorithm to tackle the non-convexity of $R_{s,kj}$.
Specifically, we first define $\mathbf{W}_k = \mathbf{w}_{k}\mathbf{w}_{k}^H$, $\mathbf{Q} = \mathbf{q}_{AN}\mathbf{q}_{AN}^H$ and $\mathbf{U} = \mathbf{u}\mathbf{u}^H$,
then the SR constraints (\ref{Rs_const}) can be equivalently rewritten as
\begin{align}
\label{Rs_log}
&f_1(\mathbf{W}_k,\mathbf{Q},\mathbf{U})+f_4(\mathbf{W}_k,\mathbf{Q},\mathbf{U})\\\nonumber
&-f_2(\mathbf{W}_k,\mathbf{Q},\mathbf{U})-f_3(\mathbf{W}_k,\mathbf{Q},\mathbf{U})\geq \gamma_s,~~\forall k,j,l,
\end{align}
where
\begin{align}
&f_1(\mathbf{W}_k,\mathbf{Q},\mathbf{U}) = \log(\sum_{k= 1}^{K}(\mathrm{Tr}(\mathbf{H}_{kj}^H\mathbf{U}\mathbf{H}_{kj}\mathbf{W}_k))\nonumber\\
&~~~~~~~~~~~~~~~~~~~~~~~~+\mathrm{Tr}(\mathbf{H}_{kj}^H\mathbf{U}\mathbf{H}_{kj}\mathbf{Q})+\sigma^2),
\end{align}
\begin{align}
&f_2(\mathbf{W}_k,\mathbf{Q},\mathbf{U}) = \log(\sum_{g\neq k}^{K}(\mathrm{Tr}(\mathbf{H}_{kj}^H\mathbf{U}\mathbf{H}_{kj}\mathbf{W}_g))\nonumber\\
&~~~~~~~~~~~~~~~~~~~~~~~~+\mathrm{Tr}(\mathbf{H}_{kj}^H\mathbf{U}\mathbf{H}_{kj}\mathbf{Q})+\sigma^2),
\end{align}
\begin{align}
&f_3(\mathbf{W}_k,\mathbf{Q},\mathbf{U}) = \log(\sum_{k = 1}^{K}(\mathrm{Tr}(\mathbf{H}_{l}^H\mathbf{U}\mathbf{H}_{l}\mathbf{W}_k))\nonumber\\
&~~~~~~~~~~~~~~~~~~~~~~~~+\mathrm{Tr}(\mathbf{H}_{l}^H\mathbf{U}\mathbf{H}_{l}\mathbf{Q})+\sigma^2),
\end{align}
\begin{align}
&f_4(\mathbf{W}_k,\mathbf{Q},\mathbf{U}) = \log(\sum_{g\neq k}^{K}(\mathrm{Tr}(\mathbf{H}_{l}^H\mathbf{U}\mathbf{H}_{l}\mathbf{W}_g)\nonumber\\
&~~~~~~~~~~~~~~~~~~~~~~~~+\mathrm{Tr}(\mathbf{H}_{l}^H\mathbf{U}\mathbf{H}_{l}\mathbf{Q})+\sigma^2).
\end{align}

Afterwards, by replacing (\ref{Rs_const}) with (\ref{Rs_log}) and dropping the rank-one constraints, the SDR of $\mathrm{(P1)}$ can be expressed as
\begin{subequations}\label{P3-1}
\begin{align}
&\text{(P2):}\min_{\{\mathbf{W}_k\},\mathbf{Q},\mathbf{U}}~ \sum_{k=1}^{K} \mathrm{Tr}(\mathbf{W}_k)+\mathrm{Tr}(\mathbf{Q})\\
&~~\text{s. t.}~~ f_1+f_4-f_2-f_3 \geq \gamma_s, ~~\forall k,j,l,\label{SR_f}\\
&~~~~~~~~\mathbf{U}_{n,n} = 1,~~\forall n=1\cdots N+1,\\
&~~~~~~~~\mathbf{U}\succeq 0, \mathbf{W}_k\succeq 0, \mathbf{Q}\succeq 0.
\end{align}
\end{subequations}

However, problem (P2) is still non-convex because constraints (\ref{SR_f}) are non-convex as well as variables $\mathbf{W}_k$ and $\mathbf{Q}$ are coupled with $\mathbf{U}$. As a result, in the following, problem (P2) is first decomposed into two non-convex subproblems. Then both the non-convex subproblems are converted into convex ones by applying the MM algorithm.
\subsection{Optimization with respect to $\mathbf{W}_k$ and $\mathbf{Q}$}
By fixing the phase shift matrix $\mathbf{U}$ as $\mathbf{U}^t$, $f_2(\mathbf{W}_k,\mathbf{Q},\mathbf{U}^t)$ and $f_3(\mathbf{W}_k,\mathbf{Q},\mathbf{U}^t)$ are concave
functions with respect to $\mathbf{W}_k$ and $\mathbf{Q}$. As a
 result, $f_2$ and $f_3$ can be upperbounded as
\begin{align}
&f_2(\mathbf{W}_k,\mathbf{Q},\mathbf{U}^t) \leq f_2(\mathbf{\tilde{W}}_k,\mathbf{\tilde{Q}},\mathbf{U}^t) \nonumber\\
&+\mathrm{Tr}(\nabla_{\mathbf{W}_k} f_2(\mathbf{\tilde{W}}_k,\mathbf{\tilde{Q}},\mathbf{U}^t)^H(\mathbf{{W}}_k-\mathbf{\tilde{W}}_k))\nonumber\\
&+\mathrm{Tr}(\nabla_{\mathbf{Q}} f_2(\mathbf{\tilde{W}}_k,\mathbf{\tilde{Q}},\mathbf{U}^t)^H(\mathbf{Q}-\mathbf{\tilde{Q}}))\nonumber\\
&\triangleq \tilde{f}_2(\mathbf{W}_k,\mathbf{Q},\mathbf{U}^t), \\
\end{align}
\begin{align}
&f_3(\mathbf{W}_k,\mathbf{Q},\mathbf{U}^t) \leq f_3(\mathbf{\tilde{W}}_k,\mathbf{\tilde{Q}},\mathbf{U}^t) \nonumber\\
&+ \mathrm{Tr}(\nabla_{\mathbf{W}_k} f_3(\mathbf{\tilde{W}}_k,\mathbf{\tilde{Q}},\mathbf{U}^t)^H(\mathbf{{W}}_k-\mathbf{\tilde{W}}_k))\nonumber\\
&+\mathrm{Tr}(\nabla_{\mathbf{Q}_{AN}} f_3(\mathbf{\tilde{W}}_k,\mathbf{\tilde{Q}},\mathbf{U}^t)^H(\mathbf{Q}-\mathbf{\tilde{Q}}))\nonumber\\
&\triangleq \tilde{f}_3(\mathbf{W}_k,\mathbf{Q},\mathbf{U}^t),
\end{align}
where
\begin{align}
&\nabla_{\mathbf{W}_k} f_2(\mathbf{{W}}_k,\mathbf{{Q}}_{AN},\mathbf{U})
=\nabla_{\mathbf{Q}} f_2(\mathbf{{W}}_k,\mathbf{{Q}}_{AN},\mathbf{U})\nonumber\\
&= \frac{1}{\mathrm{ln}2}\frac{\mathbf{H}_{kj}^H\mathbf{U}^t\mathbf{H}_{kj}}{\sum_{g\neq k}^{K}(\mathrm{Tr}(\mathbf{H}_{kj}^H\mathbf{U}^t\mathbf{H}_{kj}\mathbf{W}_g))
+\mathrm{Tr}(\mathbf{H}_{kj}^H\mathbf{U}^t\mathbf{H}_{kj}\mathbf{Q})+\sigma^2},
\end{align}
\begin{align}
&\nabla_{\mathbf{W}_k} f_3(\mathbf{{W}}_k,\mathbf{{Q}}_{AN},\mathbf{U})
=\nabla_{\mathbf{Q}} f_3(\mathbf{{W}}_k,\mathbf{{Q}}_{AN},\mathbf{U})\nonumber\\
&= \frac{1}{\mathrm{ln}2}\frac{\mathbf{H}_{l}^H\mathbf{U}\mathbf{H}_{l}}{\sum_{k = 1}^{K}(\mathrm{Tr}(\mathbf{H}_{l}^H\mathbf{U}\mathbf{H}_{l}\mathbf{W}_k))
+\mathrm{Tr}(\mathbf{H}_{l}^H\mathbf{U}\mathbf{H}_{l}\mathbf{Q})+\sigma^2}.
\end{align}

Based on the above transformation, the transmit precoder matrices $\mathbf{W}_k$ and AN matrix $\mathbf{Q}$ can be optimized by solving the following problem
\begin{subequations}\label{P2-1}
\begin{align}
&\text{(P2-1):}\min_{\{\mathbf{W}_k\},\mathbf{Q}}~ \sum_{k=1}^{K} \mathrm{Tr}(\mathbf{W}_k)+\mathrm{Tr}(\mathbf{Q})\\
&~~\text{s. t.}~~ f_1+f_4-\tilde{f}_2-\tilde{f}_3 \geq \gamma_s,~~\forall k,j,l,\\
&~~~~~~~~\mathbf{W}_k\succeq 0, \mathbf{Q}\succeq 0.
\end{align}
\end{subequations}

It can be verified that problem (P2-1) is a convex problem and can be solved by the existing solvers such as CVX \cite{CVX_cite}.
\subsection{Optimization with respect to $\mathbf{U}$}
Similarly, by fixing \{$\mathbf{W}_k$, $\mathbf{Q}$\} as \{$\mathbf{W}_k^t$, $\mathbf{Q}^t$\}, $f_2(\mathbf{W}_k^t,\mathbf{Q}_{AN}^t,\mathbf{U})$ and $f_3(\mathbf{W}_k^t,\mathbf{Q}_{AN}^t,\mathbf{U})$ are concave functions with respect to $\mathbf{U}$. Consequently, the upper-bounds of $f_2$ and $f_3$ with respect $\mathbf{U}$ can be respectively expressed as
\begin{align}
&f_2(\mathbf{W}_k^t,\mathbf{Q}^t,\mathbf{U}) \leq f_2(\mathbf{\tilde{W}}_k^t,\mathbf{\tilde{Q}}^t,\mathbf{\tilde{U}}) \nonumber\\
&+ \mathrm{Tr}(\nabla_{\mathbf{U}} f_2(\mathbf{{W}}_k^t,\mathbf{{Q}}^t,\mathbf{\tilde{U}})^H(\mathbf{{U}}-\mathbf{\tilde{U}}))
\triangleq \bar{f}_2(\mathbf{W}_k^t,\mathbf{Q}^t,\mathbf{U}), \label{f2u}\\
&f_3(\mathbf{W}_k^t,\mathbf{Q}^t,\mathbf{U}) \leq f_3(\mathbf{\tilde{W}}_k^t,\mathbf{\tilde{Q}}^t,\mathbf{\tilde{U}}) \nonumber\\
&+ \mathrm{Tr}(\nabla_{\mathbf{U}} f_3(\mathbf{{W}}_k^t,\mathbf{{Q}}^t,\mathbf{\tilde{U}})^H(\mathbf{{U}}-\mathbf{\tilde{U}}))
\triangleq \bar{f}_3(\mathbf{W}_k^t,\mathbf{Q}^t,\mathbf{U}) \label{f3u},
\end{align}
where
\begin{align}
&\nabla_{\mathbf{U}} f_2(\mathbf{W}_k^t,\mathbf{Q}_{AN}^t,\mathbf{U})\nonumber\\
&= \frac{1}{\mathrm{ln}2}\frac{\sum_{g\neq k}^{K}(\mathbf{H}_{kj}\mathbf{W}_g^t\mathbf{H}_{kj}^H)
+\mathbf{H}_{kj}\mathbf{Q}^t\mathbf{H}_{kj}^H}{\sum_{g\neq k}^{K}(\mathrm{Tr}(\mathbf{H}_{kj}^H\mathbf{U}\mathbf{H}_{kj}\mathbf{W}_g^t))
+\mathrm{Tr}(\mathbf{H}_{kj}^H\mathbf{U}\mathbf{H}_{kj}\mathbf{Q}^t)+\sigma^2},
\end{align}
\begin{align}
&\nabla_{\mathbf{U}} f_2(\mathbf{W}_k^t,\mathbf{Q}^t,\mathbf{U})\nonumber\\
&=\frac{1}{\mathrm{ln}2}\frac{\sum_{k=1}^{K}(\mathbf{H}_{l}\mathbf{W}_g^t\mathbf{H}_{l}^H)
+\mathbf{H}_{l}\mathbf{Q}^t\mathbf{H}_{l}^H}{\sum_{k = 1}^{K}(\mathrm{Tr}(\mathbf{H}_{l}^H\mathbf{U}\mathbf{H}_{l}\mathbf{W}_k))
+\mathrm{Tr}(\mathbf{H}_{l}^H\mathbf{U}\mathbf{H}_{l}\mathbf{Q})+\sigma^2}.
\end{align}

According to \eqref{f2u} and \eqref{f3u}, the optimization problem of the phase shift matrix $\mathbf{U}$ is given by
\begin{subequations}\label{P2-2}
\begin{align}
&\text{(P2-2): } \text{find}~{\mathbf{U}}\\
&~~\text{s. t.}~~ f_1+f_4-\bar{f}_2-\bar{f}_3 \geq \gamma_s,~~\forall~k,j,l,\\
&~~~~~~~~\mathbf{U}_{n,n} = 1,~~\forall n=1\cdots N+1,\\
&~~~~~~~~\mathbf{U}\succeq 0,
\end{align}
\end{subequations}

Problem (P2-2) is a standard convex optimization problem, which can be optimally solved existing convex
optimization solvers (e.g., CVX) \cite{CVX_cite}.
\subsection{Overall Algorithm and Complexity Analysis}
By optimizing the problem (P2-1) and problem (P2-2) alternately, we obtain a sub-optimal solution to problem (P2).
However, the solution obtained by solving problem (P2) can not be guaranteed to be a feasible solution of the original problem (P1) since the rank-one constraints are relaxed in problem (P2). To address this problem, the Gaussian randomization method is used to recover the rank-one solutions.
Different from \cite{wu_one} that the Gaussian randomization method is used in each iteration, we apply the Gaussian randomization method only once when obtaining the final solution of problem (P2). This is due to the fact using the Gaussian randomization method in each iteration may lead to non-convergence and high complexity. The SDR-based AO algorithm is summarized in Algorithm 1.
\begin{algorithm}
\begin{enumerate}
  \item \textbf{Initialization:} $\mathbf{W}_k$ and $\mathbf{Q},\mathbf{U}$, convergence accuracy $\epsilon$ and set $t = 0$.
  \item \textbf{repeat}
  \item ~~~~Set t = t + 1.
  \item ~~~~Fix $\mathbf{U} = \mathbf{U} ^t$ update $\mathbf{W}_k$ and $\mathbf{Q}$ with $\mathbf{W}_k^{t+1}$ and $\mathbf{Q}^{t+1}$by solving problem $\mathrm{(P2-1)}$.
  \item ~~~~Fix $\mathbf{W}_k = \mathbf{W}_k^{t+1}$ and $\mathbf{Q}=\mathbf{Q}^{t+1}$ , calculate the $\mathbf{U}^{t+1}$ by solving problem $\mathrm{(P2-2)}$.
  \item \textbf{until} $|P_{t+1}-P_t|<\epsilon$.
   \item\textbf{Gaussian randomization}.
\end{enumerate}
\caption{The SDR-based Alternating Optimization Algorithm }\label{algorithm 1}
\end{algorithm}

Because the objective value of problem (P2) of the proposed SDR-based method decrease in each iteration. Besides, the optimal value of (P2) has a lower bound due to the SR constraints. Therefore, the convergence of the proposed SDR-based AO algorithm can be guaranteed.

In the following, we analyze the computational complexity of Algorithm 1. Observing that problem (P2-1) has $TL$ LMI constraints of size 1, $K$ LMI constraints of size $M$ and 1 LMI constraints of size $M$. The number of decision variables $n_1=(K+1)M^2$. Problem (P2-2) has $TL$ LMI constraints of size 1, 1 LMI constraints of size 1, 1 LMI constraints of size $N$. The number of decision variables $n_2=N^2$ \cite{complexity}. Hence, the overall complexity of Algorithm 1 is
\begin{align}
&\mathcal{O}\Big(n_1D_1\sqrt{TL+KM+M}\big((TL+KM^3+M^3)\nonumber\\
&+n_1(TL+KM^2+M^2)+n_1^2\big)\nonumber\\
&+n_2D_2\sqrt{TL+1+N}\big((TL+N^3+1)\nonumber\\
&+n_2(TL+N^2+1)+n_2^2\big)\Big),
\end{align}
where $D_1$ and $D_2$  denote the numbers of iterations in problem (P2-1) and in problem (P2-2), respectively.
It is observed that the computational complexity is on the order of $M^{8.5}$ or $N^{8.5}$, which is externely high and unpractical. Hence, we will propose a low complexity SOCP-based algorithm in the next section.
\section{Low-complexity SOCP-based Algorithm}
In this section, we aim to propose a low-complexity but efficient SOCP-based algorithm to solve problem (P1).

Define $\gamma_{b(k,j)}$ and $\gamma_{e(k,l)}$ as the minimum signal to interference plus noise ratio (SINR) at the $j$-th user in the $k$-th Bob group and the maximum SINR at the $l$-th Eve, respectively. Then, the optimization problem (P1) can be mathematically recast as
\begin{subequations}\label{P2}
\begin{align}
\text{(P3):}&\min_{\{\mathbf{w}_k\},\mathbf{q}_{AN},\mathbf{v},\bm{\gamma}_{b},\bm{\gamma}_{e}}~~~ \sum_{k=1}^{K}\|\mathbf{w}_k\|^2+\|\mathbf{q}_{AN}\|^2\\
&\text{s. t.}~~~~\mathrm{SINR}_{b(k,j)}\geq \gamma_{b(k,j)},~~\forall k,j, \label{sinr_rb}\\
&~~~~~~~~\mathrm{SINR}_{e(k,l)}\leq \gamma_{e(k,l)},~~\forall k,l, \label{sinr_re}\\
&~~~~~~~~{1+\gamma_{b(k,j)}}\geq {2^{\gamma_s}}(1+\gamma_{e(k,l)})~~\forall k,j,l,\label{RS_rb_re}\\\label{v_unit}
&~~~~~~~~|v_n|^2 = 1,~~\forall n=1\cdots N.
\end{align}
\end{subequations}

Note that the optimization variables \{$\mathbf{w}_k$, $\mathbf{q}_{AN}$\} and $\mathbf{v}$ are mutually coupled in constraints (\ref{sinr_rb}) and (\ref{sinr_re}). Moreover, the constraints (\ref{v_unit}) are uni-modular. Hence, it is non-trivial to solve this problem. In the following, we optimize problem (P3) by applying alterative manner.
\subsection{Optimization with respect to $\mathbf{w}_k$ and $\mathbf{q}_{AN}$}
For given phase shifts at the IRS, the problem (P3) is reduced to
\begin{subequations}\label{P-1}
\begin{align}
&\text{(P3-1):}\min_{\{\mathbf{w}_k\},\mathbf{q}_{AN},\bm{\gamma}_{b},\bm{\gamma}_{e}}~~~ \sum_{k=1}^{K}\|\mathbf{w}_k\|^2+\|\mathbf{q}_{AN}\|^2\\
&\text{s. t.}~~~~\mathrm{SINR}_{b(k,j)}(\mathbf{w}_k,\mathbf{q}_{AN})\geq \gamma_{b(k,j)},~~\forall k,j,\label{SINR_B}\\
&~~~~~~~~\mathrm{SINR}_{e(k,l)}(\mathbf{w}_k,\mathbf{q}_{AN})\leq \gamma_{e(k,l)},~~\forall k,l,\label{SINR_E}\\
&~~~~~~~~(\ref{RS_rb_re}),
\end{align}
\end{subequations}

Problem (P3-1) is non-convex due to the non-convex constraints (\ref{SINR_B}) and (\ref{SINR_E}). To address them, we focus on converting them into convex ones.

Note that (\ref{SINR_B}) and (\ref{SINR_E}) can be respectively rearranged as
\begin{align}
&{\sum_{g\neq k}^{K}|\mathbf{u}^H\mathbf{H}_{kj}\mathbf{w}_g|^2
+|\mathbf{u}^H\mathbf{H}_{kj}\mathbf{q}_{AN}|^2+\sigma^2} \label{SINR_B1}\\\nonumber
&\leq \frac{|\mathbf{u}^H\mathbf{H}_{kj}\mathbf{w}_k|^2}{\gamma_{b(k,j)}},~\forall k,j,\\
&\frac{|\mathbf{u}^H\mathbf{H}_{l}\mathbf{w}_k|^2}{\gamma_{e(k,l)}} \leq
{\sum_{g\neq k}^{K}|\mathbf{u}^H\mathbf{H}_{l}\mathbf{w}_g|^2+|\mathbf{u}^H\mathbf{H}_{l}
\mathbf{q}_{AN}|^2+\sigma^2} ,~~\forall k,l.\label{SINR_E1}
\end{align}
Note that constraints \eqref{SINR_B1} and \eqref{SINR_E1} are in the form of the
superlevel of convex functions, which allows us to apply the first-order approximation technique to transform them into convex constraints. Specifically, for a complex value $x$, it is well known that
\begin{align}
\frac{|x|^2}{r}\geq \frac{2\Re(\tilde{x}^*x)}{\tilde{r}}-\frac{\tilde{x}^*\tilde{x}}{\tilde{r}^2}r\triangleq F(x,r,\tilde{x},\tilde{r}). \label{appro}
\end{align}
Based on \eqref{appro}, (\ref{SINR_B1}) and (\ref{SINR_E1}) can be respectively approximated as
\begin{align}
&{\sum_{g\neq k}^{K}|\mathbf{u}^H\mathbf{H}_{kj}\mathbf{w}_g|^2
+|\mathbf{u}^H\mathbf{H}_{kj}\mathbf{q}_{AN}|^2+\sigma^2} \label{SINR_B2}\\\nonumber
&\leq F(\mathbf{u}^H\mathbf{H}_{kj}\mathbf{w}_k, \gamma_{b(k,j)},\mathbf{u}^H\mathbf{H}_{kj}\mathbf{\tilde{w}}_k ,\tilde{\gamma}_{b(k,j)}),~~\forall k,j,
\end{align}
\begin{align}
&\frac{|\mathbf{u}^H\mathbf{H}_{l}\mathbf{w}_k|^2}{\gamma_{e(k,l)}} \leq
\sum_{g\neq k}^{K}F(\mathbf{u}^H\mathbf{H}_{l}\mathbf{w}_g, 1,\mathbf{u}^H\mathbf{H}_{l}\mathbf{\tilde{w}}_g ,1)\label{SINR_E2}\\\nonumber
&~~~~~+F(\mathbf{u}^H\mathbf{H}_{l}
\mathbf{q}_{AN}, 1,\mathbf{u}^H\mathbf{H}_{l}
\mathbf{\tilde{q}}_{AN} ,1)+\sigma^2,~~\forall k,l.
\end{align}

According to the above transformation, problem (P3-1) can be converted into the following problem
\begin{subequations}\label{P3-1}
\begin{align}
\text{(P3-1'):}&\min_{\{\mathbf{w}_k\},\mathbf{q}_{AN},\bm{\gamma}_{b},\bm{\gamma}_{e}}~~~ \sum_{k=1}^{K}\|\mathbf{w}_k\|^2+\|\mathbf{q}_{AN}\|^2\\
&\text{s. t.}~~~~(\ref{SINR_B2}),(\ref{SINR_E2}),(\ref{RS_rb_re}).
\end{align}
\end{subequations}
Problem (P3-1') is an SOCP problem and  its optimal solution can be found by using CVX.
\subsection{Optimization with respect to $\mathbf{v}$}
By fixing $\mathbf{w}_k$ and $\mathbf{q}_{AN}$, problem (P3) can be reduced to
\begin{subequations}\label{P3-2}
\begin{align}
\text{(P3-2):}&\min_{\mathbf{u},\bm{\gamma}_{b},\bm{\gamma}_{e}}~~~ 1\\
&\text{s. t.}~~~~\mathrm{SINR}_{b(k,j)}(\mathbf{u})\geq \gamma_{b(k,j)},~~\forall k,j,\label{SINRB_V}\\
&~~~~~~~~\mathrm{SINR}_{e(k,l)}(\mathbf{u})\leq \gamma_{e(k,l)},~~\forall k,l,\label{SINRE_V}\\
&~~~~~~~~(\ref{RS_rb_re}),\\\label{UNI-MO}
&~~~~~~~~|u_n| = 1,~~\forall n=1\cdots N, u_{N+1} = 1.
\end{align}
\end{subequations}
It is observed that problem (P3-2) is still non-convex due to the non-convex constraints (\ref{SINRB_V}) and (\ref{SINRE_V}) as well as the unit-modulus constraint (\ref{UNI-MO}), which leads to problem (P3-2) difficult to solve. Therefore, we concentrate on dealing with these constraints in the next.

For the non-convex constraints (\ref{SINRB_V}) and (\ref{SINRE_V}), similar to (\ref{SINR_B1}) and (\ref{SINR_E1}), they can be transformed into
\begin{align}
&{\sum_{g\neq k}^{K}|\mathbf{u}^H\mathbf{H}_{kj}\mathbf{w}_g|^2
+|\mathbf{u}^H\mathbf{H}_{kj}\mathbf{q}_{AN}|^2+\sigma^2} \label{SINR_VB2}\\\nonumber
&\leq F(\mathbf{u}^H\mathbf{H}_{kj}\mathbf{w}_k, \gamma_{b(k,j)},\mathbf{\tilde{u}}^H\mathbf{H}_{kj}\mathbf{{w}}_k ,\tilde{\gamma}_{b(k,j)}),~~\forall k,j,\\
&\frac{|\mathbf{u}^H\mathbf{H}_{l}\mathbf{w}_k|^2}{\gamma_{e(k,l)}} \leq
\sum_{g\neq k}^{K}F(\mathbf{u}^H\mathbf{H}_{l}\mathbf{w}_g, 1,\mathbf{\tilde{u}}^H\mathbf{H}_{l}\mathbf{{w}}_g ,1)\label{SINR_VE2}\\\nonumber
&~~~~~+F(\mathbf{\tilde{u}}^H\mathbf{H}_{l}
\mathbf{q}_{AN}, 1,\mathbf{u}^H\mathbf{H}_{l}
\mathbf{{q}}_{AN} ,1)+\sigma^2,~~\forall k,l,
\end{align}
respectively. So far, the non-convex constraints (\ref{SINRB_V}) and (\ref{SINRE_V}) have been converted into the convex constraints \eqref{SINR_VB2} and \eqref{SINR_VE2}, respectively.

For the unit-modulus constraint (\ref{UNI-MO}), it can be relaxed as
\begin{align}
\label{relax_uni}
|u_n|^2 \leq 1,~~\forall n=1\cdots N, u_{N+1} = 1,
\end{align}
which is convex now.

Following the above transformation, problem (P3-2) can be recast as the following SOCP problem
\begin{subequations}\label{P3-1}
\begin{align}
\text{(P3-2'):}&\min_{\mathbf{u},\bm{\gamma}_{b},\bm{\gamma}_{e}}~~~ 1\\
&\text{s. t.}~~~~(\ref{SINR_VB2}),(\ref{SINR_VE2}),(\ref{relax_uni}),
\end{align}
\end{subequations}
which can be efficiently solved by CVX.
Denote by $\{\mathbf{u}^\dagger, \bm{\gamma}_{b}^*,\bm{\gamma}_{e}^*\}$ the optimal solution to problem (P3-2'). Then the solution to problem (P3-2) can be expressed as $\{\mathbf{u}^{*}, \bm{\gamma}_{b}^*,\bm{\gamma}_{e}^*\}$, where
\begin{align}
\label{relax_uni}
\mathbf{u}^{*} = e^{j\angle{\frac{\mathbf{u}^\dagger}{\mathbf{u}_{N+1}^\dagger}}}.
\end{align}
\subsection{Overall Algorithm and Complexity Analysis}
Based on the above two subsections, the low-complexity SOCP-based algorithm is summarized in Algorithm 2.
\begin{algorithm}
\begin{enumerate}
  \item \textbf{Initialization:} $\mathbf{w}_k$ and $\mathbf{q},\mathbf{u}$, convergence accuracy $\epsilon$ and set $r = 0$.
  \item \textbf{repeat}
  \item ~~~~Set $r = r + 1$.
  \item ~~~~With given $\mathbf{u} = \mathbf{u} ^r$ update $\mathbf{w}_k$ and $\mathbf{q}_{AN}$ with $\mathbf{w}_k^{t+1}$ and $\mathbf{q}_{AN}^{r+1}$by solving problem $\mathrm{(P3-1')}$.
  \item ~~~~Fix $\mathbf{w}_k = \mathbf{w}_k^{r+1}$ and $\mathbf{q}_{AN}=\mathbf{q}_{AN}^{r+1}$ , calculate the $\mathbf{u}^{r+1}$ by solving problem $\mathrm{(P3-2')}$ and $(\ref{relax_uni})$.
  \item \textbf{until} $|P_{r+1}-P_r|<\epsilon$.
\end{enumerate}
\caption{The Low-complexity SOCP-based Algorithm }\label{algorithm 2}
\end{algorithm}

 Due to the fact that the objective value of problem (P3) is non-increasing in each iteration. Meanwhile, the objective value of problem (P3) is lower bounded by a finite value, thus the convergence of Algorithm 2 can be guaranteed.

In terms of the complexity, problem (P3-1') consists of $T$ SOC constraints of dimension $K+3$, $KL$ SOC of dimension $3$ and $TL$ LMI of size 1. The number of decision variables $n_1=M+M+T+KL$. Problem (P3-2') includes $T$ SOC constraints of dimension $K+3$, $KL$ SOC constraints of dimension $3$ and $TL+N+1$ LMI of size 1. The number of decision variables $n_2=N+T+KL$. Therefore, the computation complexity is
\begin{align}
&\mathcal{O}\Big(n_1D_1\sqrt{TL+2(T+KL)}\nonumber\\
&\big(T(K+3)^2+3^2KL+n_1(TL+1)+n_1^2\big)\nonumber\\
&+n_2D_2\sqrt{TL+N+1+2(T+KL)}\nonumber\\
&\big(T(K+3)^2+3^2KL+n_2(TL+N+1)+n_2^2\big)\Big),
\end{align}
where $D_1$ and $D_2$  denote the numbers of iterations in problem (P3-1') and problem (P3-2'), respectively.
Obviously, computational complexity is on the order of $M^{3}$ or $N^{3}$, which is much lower that of Algorithm 1 (i.e., $M^{8.5}$ or $N^{8.5}$).

\section{Simulation and analysis}
\begin{figure}[htb]
  \centering
  \includegraphics[width=0.48\textwidth]{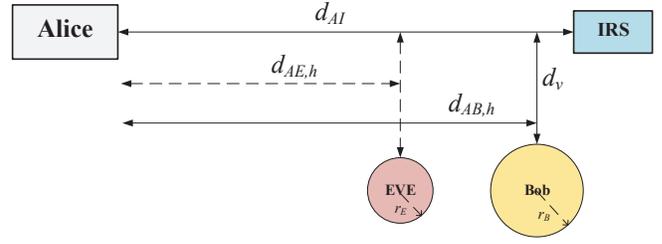}
  \caption{Simulation setup.}
  \label{Simu_Mod}
\end{figure}

In this section, numerical results are provided to evaluate the performance of our proposed two algorithms. In this paper, we consider a system as shown in Fig.~\ref{Simu_Mod}, where $d_{AI}=70$ m, $d_{AE,h} = 60$ m, $d_{AB,h} = 70$ m, $d_v = 5$ m, $r_E=2.5$ m, and $r_B = 5$ m. All the channels are assumed to follow the Rayleigh fading model and the path loss at the distance $d$ is modeled as $PL(d) = PL_0-10\alpha \log_{10}(\frac{d}{d_0})$, where $PL_0 = -30$ dB denotes the path loss at the reference distance $d_0 = 1$ m, $\alpha$ denotes the path loss exponent. Specifically, the path loss exponents of the Alice-IRS, IRS-Eves/Bobs and Alice-Eves/Bobs channels are set to be 2.2, 2.5 and 3.5, respectively. The other simulation parameters are set as: $\sigma_b^2 = \sigma_e^2 = -90$ dBm, $K=2$, $|\mathcal{G}_1| = |\mathcal{G}_2| = 2$ , $L = 2$. For comparision, the benchmark schemes are given as follows:
\begin{itemize}
  \item Without IRS: There is no use of the IRS and only the transmit beamformer $\mathbf{w}_k$ and AN are designed \cite{Sec_MuG}.
  \item Random phase shifts: The phase shifts of IRS are set randomly in $[0,2\pi]$, the transmit beamformer $\mathbf{w}_k$ and AN are optimized.
\end{itemize}

\begin{figure}[htb]
  \centering
  \includegraphics[width=0.48\textwidth]{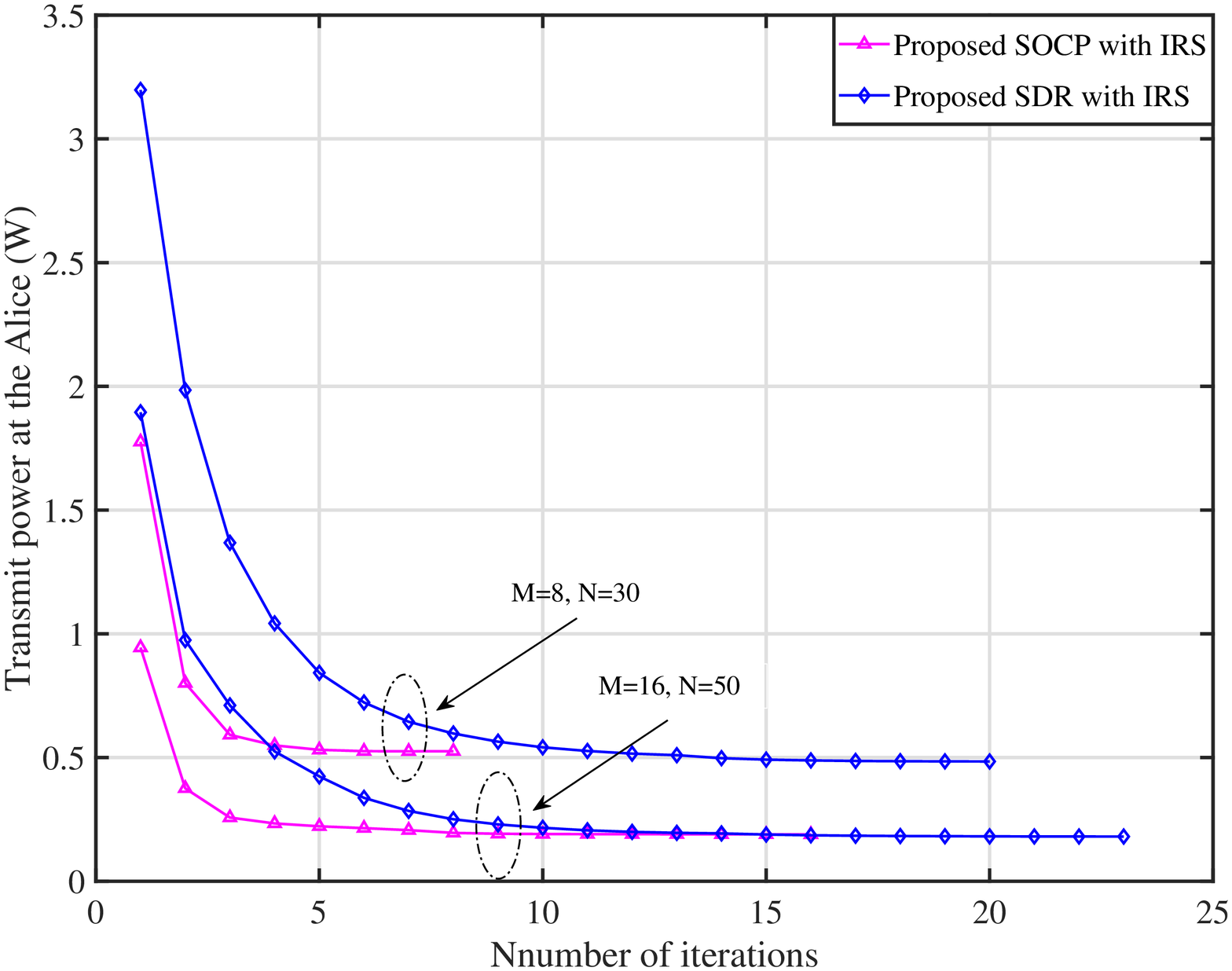}
  \caption{Convergence performance of different algorithms.}
  \label{convergence}
\end{figure}

Fig.~\ref{convergence} demonstrates the convergence performance of the proposed SDR and SCOP algorithms with $\gamma_s$ = 2 bps/Hz. It is observed that the both algorithms converge within a small number of iterations under different setups. Besides, it is worth noticing that the proposed SOCP-based algorithm not only has less computation complexity than the proposed SDR-based algorithm, but also converges faster.

\begin{figure}[htb]
  \centering
  \includegraphics[width=0.48\textwidth]{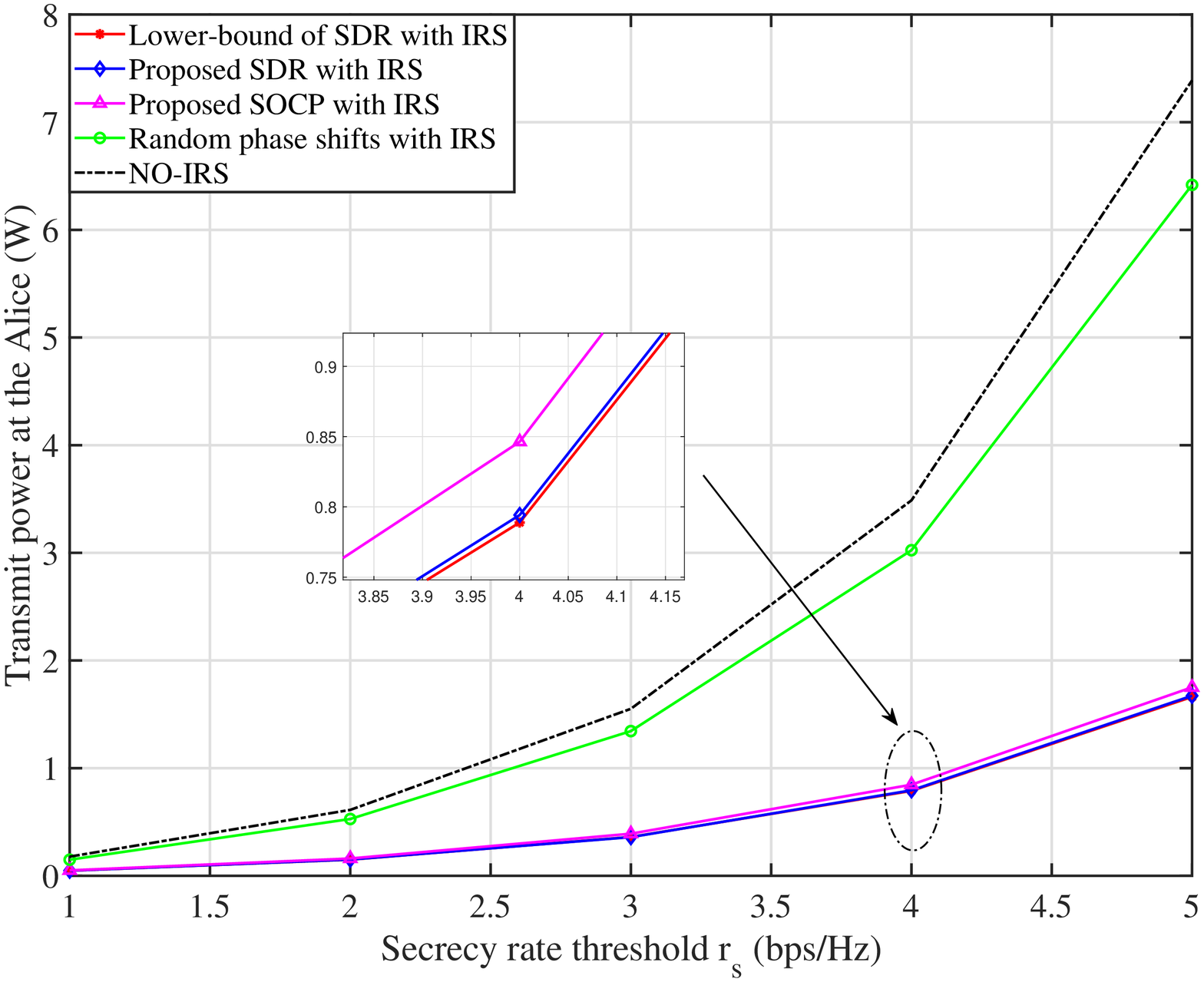}
  \caption{Transmit power at the Alice versus secrecy rate threshold $\gamma_s$.}
  \label{RS}
\end{figure}

Fig.~\ref{RS} shows the minimum required transmit power at the Alice under different values of SR threshold $r_s$ with $M=8$ and $N=50$. From Fig.~\ref{RS}, we can first notice that, as expected, the minimum required transmit power at the Alice obtained by all the schemes increase as the $\gamma_s$ increases.
Then, it is observed that the both proposed algorithms have the similar performance. This is due to the fact that the both algorithms could guarantee to converge to a local or even global optimum. Finally, we can note that the proposed schemes outperform the NO-IRS scheme as well as the random phase shifts scheme, and the  performance gap increases with $\gamma_s$. The former validates the advantages of the IRS in the multigroup multicast system, and the later presents the effectiveness of optimizing the phase shifts at the IRS.

\begin{figure}[htb]
  \centering
  \includegraphics[width=0.48\textwidth]{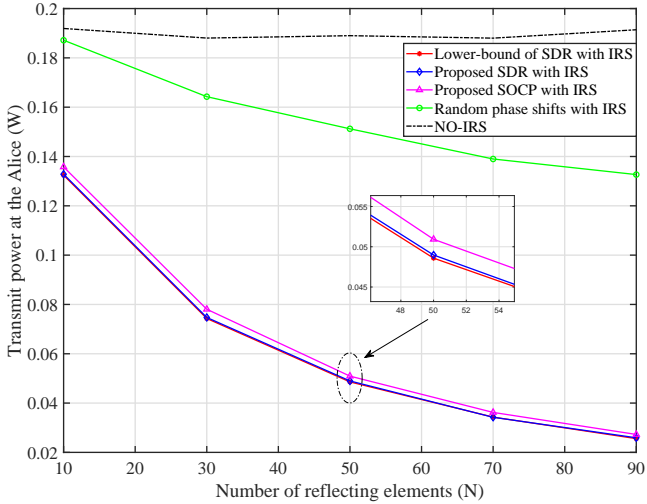}
  \caption{Transmit power at the Alice versus the number of reflecting elements $N$.}
  \label{P-N}
\end{figure}

The required transmit power at the Alice versus the number of reflecting elements at the IRS $N$ is shown in Fig.~\ref{P-N} with $M=8$ and $\gamma_s$ = 1 bps/Hz. It is noted that the transmit power at the Alice obtained by the both proposed schemes decreases significantly as $N$ decreases because the IRS located near Bobs and the received signal at Bobs mainly comes from the reflection of the IRS. Besides, it is worth noting that the performance gap between the IRS-aided schemes and the NO-IRS scheme increases as $N$ increases. This is expected since the proportion of the signal reflected by the IRS to the received signal at Bobs increases as $N$ increases.
Additionally, when $N$ is relatively small, we also find that the two proposed methods require almost only half of the transmit power compare to the case without IRS. This further verify that the introduction of IRS is useful to improve system performance.

\begin{figure}[htb]
  \centering
  \includegraphics[width=0.48\textwidth]{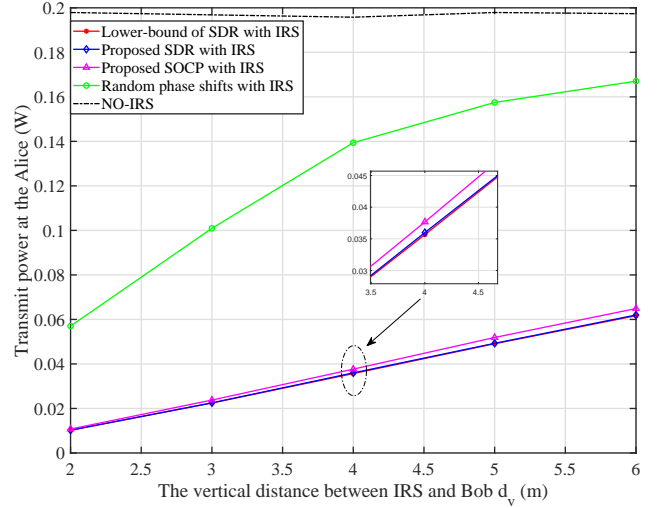}
  \caption{Transmit power at the Alice versus the vertical distance $d_v$.}
  \label{P-dv}
\end{figure}

Fig.~\ref{P-dv} shows the required transmit power at the Alice versus the vertical distance $d_v$ with $M=8$, $N = 50$ and $\gamma_s$ = 1 bps/Hz. It is observed that the required transmit power obtained by the IRS-aided schemes increases as $d_v$ increases. This is due to the fact that the larger $d_v$, the farther distances between the IRS and the Bobs as well as the Eves. As a result, more transmit power is required to meet the SR constraints. It is worth noting that the transmit power obtained by the NO-IRS scheme almost remains unchanged. This is because the change of $d_v$ has little effect on the distances between the Alice and the Bobs as well as the Eves when $d_{AB,h}$ and $d_{AE,h}$ are much greater than $d_v$. Note that the transmit power require by using the random phase shifts at the IRS is much higher than that required by using the two proposed schemes. This verify that it is necessary to optimize the phase shifts at the IRS and the proposed algorithms are effective to reduce the transmit power.

\begin{figure}[htb]
  \centering
  \includegraphics[width=0.48\textwidth]{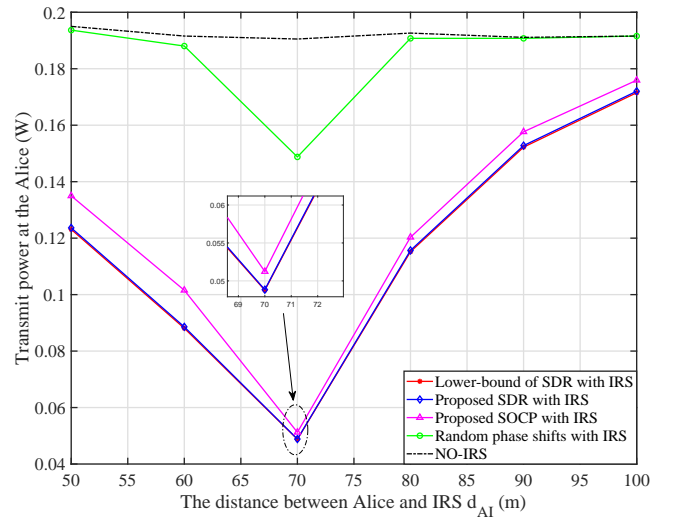}
  \caption{Transmit power at the Alice versus the distance $d_{AI}$.}
  \label{P-dAI}
\end{figure}

In Fig.~\ref{P-dAI}, we study the transmit power at the Alice versus the distance between the Alice and the IRS $d_{AI}$ with $M=8$, $N = 50$ and $\gamma_s$ = 1 bps/Hz. As expected, the proposed both schemes requires less transmit power than the both benchmark schemes. In particular, when there is no IRS, the transmit power at the Alice barely changes as $d_{AI}$ increases. Besides, for the three IRS-aided schemes, we can observed that the transmit power decreases  with the increase of $d_{AI}$ when $d_{AI}\leq 70$ m, and the phenomenon reverses when $d_{AI} > 70$ m. This is because the distance between the IRS and the Bobs decreases as $d_{AI}$ increases when $d_{AI}\leq 70$ m, while the distance increases when $d_{AI} > 70$ m.

\begin{figure}[htb]
  \centering
  \includegraphics[width=0.48\textwidth]{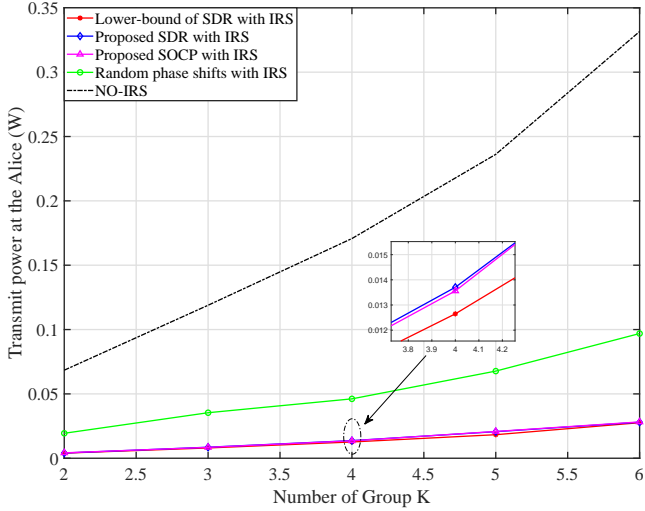}
  \caption{Transmit power at the Alice versus the number of group $K$.}
  \label{P-K}
\end{figure}

In Fig~\ref{P-K}, we gradually increase the number of group $K$ (each group has two user, which are randomly located in the Bobs cluster) to study its effect on the transmit power at the Alice with $M=8$, $N = 50$, $d_v = 2$ m and $\gamma_s$ = 0.5 bps/Hz. Obviously, the transmit power at the Alice of all the schemes increases as $K$ increases. Compare to the case without IRS, the transmit power required by applying the two proposed algorithm with IRS is significantly reduced. Additional, the performance gap between the NO-IRS scheme and the three IRS-aided schemes increases rapidly with the increase of $K$, which further demonstrates the effectiveness of the IRS on enhancing the performance of the multigroup multicast system.

\section{Conclusion}
In this paper, we investigated a novel IRS-aided secure multigroup multicast MISO communication system. By jointly optimizing the transmit beamformer, AN vector and phase shifts at the IRS, we minimized the transmit power at the Alice subject to the secrecy rate constraints. For this non-convex optimization problem, we first proposed an SDR method based on the alterative optimization and obtained a high-quality solution. Due to the high computation complexity of the proposed SDR method, the SOCP method with low complexity is then presented. The simulation results demonstrate that the proposed SOCP algorithm can obtain the similar performance as the proposed SDR algorithm. Besides, it is shown that the transmit power required at the Alice of the two proposed schemes have a significant drop compare to that of the scheme without IRS.
\ifCLASSOPTIONcaptionsoff
  \newpage
\fi
\bibliographystyle{IEEEtran}
\bibliography{IEEEfull,cite}
\end{document}